\documentclass[aps,twocolumn,amsmath,amssymb,amsfonts,longbibliography,nobibnotes]{revtex4-1} 

\usepackage[utf8]{inputenc}
\usepackage[T1]{fontenc} 

\usepackage{graphicx} 
\usepackage{sidecap}
\usepackage{times}
\usepackage{mathptmx}
\usepackage[hypertexnames=false,colorlinks=true,citecolor=blue,linkcolor=blue,urlcolor=blue]{hyperref}

\def\dyn{\mathrm{dyn}}
\def\AA{\mathrm{AA}}
\def\Berry{\mathrm{Berry}}
\def\geo{\mathrm{geo}}
\def\ex{\mathrm{ex}}

\begin{document}

\title{Majorana phase-gate based on the geometric phase}
\author{Andrzej Wi\k{e}ckowski}
\email{andrzej.wieckowski@pwr.edu.pl}
\def\PWr{Department of Theoretical Physics, 
Faculty of Fundamental Problems of Technology,
Wroc\l{}aw University of Science and Technology,  
PL-50370 Wroc\l{}aw,  Poland}
\author{Marcin Mierzejewski}
\email{marcin.mierzejewski@pwr.edu.pl}
\author{Michał Kupczyński}
\email{michal.kupczynski@pwr.edu.pl}

\affiliation{\PWr}

\begin{abstract}
We study  dynamics of a single qubit encoded in two pairs of Majorana modes, whereby each pair is hosted on a trijunction described by the Kitaev model extended by many-body interactions.
We demonstrated that the challenging phase-gate may be efficiently implemented via braiding of partially overlapping modes.  
Although such qubit acquires both geometric and dynamical  phases during the braiding protocol, the latter phase may be eliminated if the Majorana modes are hosted by systems with appropriate particle-hole symmetry. 
\end{abstract}

\maketitle

\section{Introduction}\label{sec.introduction}
The Majorana zero-energy modes (MZMs) have recently attracted a significant interest as building blocks of the future topological quantum computers~\cite{ivanov.01,sarma.freedman.05,bonderson.freedman.08,nayak.simon.08,akhmerov.10,sarma.freedman.15,plugge.landau.16,aasen.hell.16,karzig.knapp.17,agaudo.17,lutchyn.bakkers.18}.
So far, the experimental and theoretical studies have focused mostly on finding an optimal physical system that hosts the MZM~\cite{sticlet.beba.12,ptok.kobialka.17,maska.domanski.17,maska.gorczyca-goraj.17,li.jeon.18,kobialka.domanski.18,kobialka.ptok.19} 
as well as on  developing appropriate techniques which clearly confirm the existence of MZM therein~\cite{liu.sau.17,liu.sau.18,hell.flensberg.18}. 
Recent experimental results strongly support the presence of the MZM in 
superconductor--semiconductor hybrid nanostructures~\cite{deng.yu.12,mourik.zuo.12,das.ronen.12,finck.vanharlingen.13,nichele.drachmann.17,gul.zhang.18,deng.vaitiekenas.16,deng.vaitieke.18,zhang.liu.18,wang.kong.18},
in one-dimensional monoatomic chains deposited on the surface of superconductors~\cite{nadjperge.drozdov.14,pawlak.kisiel.16,feldman.randeria.16,ruby.heinrich.17,jeon.xie.17,kim.palaciomorales.18},
 in the superconducting vortices \cite{sun.jia.17,machida.sun.19,jiang.dai.19,chiu.machida.19}
and  in two-dimensional topological superconductors ~\cite{menard.guissart.17,palaciomorales.mascot.18}.

The fundamental problem for quantum computing is to effectively implement
the set of the universal gates which consists of the Hadamard gate, the Z gate and also the $\pi/8$-gate (phase-gate)~\cite{nielsen.chuang.11}. The general scheme for building the former two gates is already well established via topologically protected braiding operations of MZMs~\cite{alicea.oreg.11,heck.akhmerov.12,karzig.yu.14,wu.liang.14,pedrocchi.divincenzo.15,cheng.he.16,li.neupert.16,matos.shabani.17,sekania.plugge.17,bauer.karzig.18, ritland.rahmani.18, malciu.mazza.18}. However, the phase-gate poses a challenging problem, since the latter operations are insufficient for its implementation~\cite{sarma.freedman.15}. The very basic method of overcoming this problem is to bring two Majorana quasiparticles close to each other~\cite{sarma.freedman.15}. 
The MZMs are operators which  map an eigenstate from one parity sector to a state in another sector with identical energy. Bringing two MZMs together lifts the latter degeneracy (MZMs are no longer strict zero-modes) and splits the levels for odd and even numbers of particles by $\delta E$. 
In principle, the phase-shift needed for the phase-gate can be obtained via fine-tuning of two parameters: $\delta E$ and the period of time $\Delta t$ for which the MZMs are brought close to each other. 
However, the resulting phase is not protected by any symmetry and, as a consequence, each such operation must be followed by an error correction, e.g. via the magic state distillation~\cite{bravyi.kitaev.05}. 

The phase-shift induced via proximity of two  MZMs is a dynamical phase. Such operation requires a precise control of two independent parameters: $\delta E$ and $\Delta t$. 
In the present work we derive other possibility in which fine-tuning of $\Delta t$ is eliminated.
It consists in double braiding of two MZMs, which are previously brought together, so that the Majorana edge states partially overlap in the real space.
Braiding of such overlapping modes leads to a small shift of the {\it geometric phase} with respect to results for spatially separated MZMs~\cite{sekania.plugge.17}. 
The geometric phase  is independent of the braiding time, $\Delta t$, hence it is better suited  building of the phase-gate than simple protocol based on the dynamical phase. 
Despite an apparent advantage of such protocol, an important problem remains to be solved: qubit built out of overlapping MZMs acquires during its evolution not only the geometric but also the dynamical phase, whereby the latter occurs due to the energy splitting $\delta E$.
However, we demonstrate that the dynamical phase may be effectively eliminated if the MZMs are hosted by appropriate systems with particle--hole symmetry. 
The latter property is shown to hold  also for systems with many-body interactions.

The paper is organized as follows:
in Section~\ref{sec.model} we recall the method of storing
a qubit in 4 MZM (sparse encoding) and specify the microscopic
model of a system that hosts MZMs;
next, in Section~\ref{sec.results} we present our numerical results
concerning the geometric- and dynamical-phases gained after the braiding
of overlaping Majorana modes and show how the latter phase may be eliminated; finally, we summarize our results in Section~\ref{sec.summary}.

\section{Model and details of braiding}\label{sec.model}
\begin{figure}[t]
    \centering
    \includegraphics[width=\columnwidth]{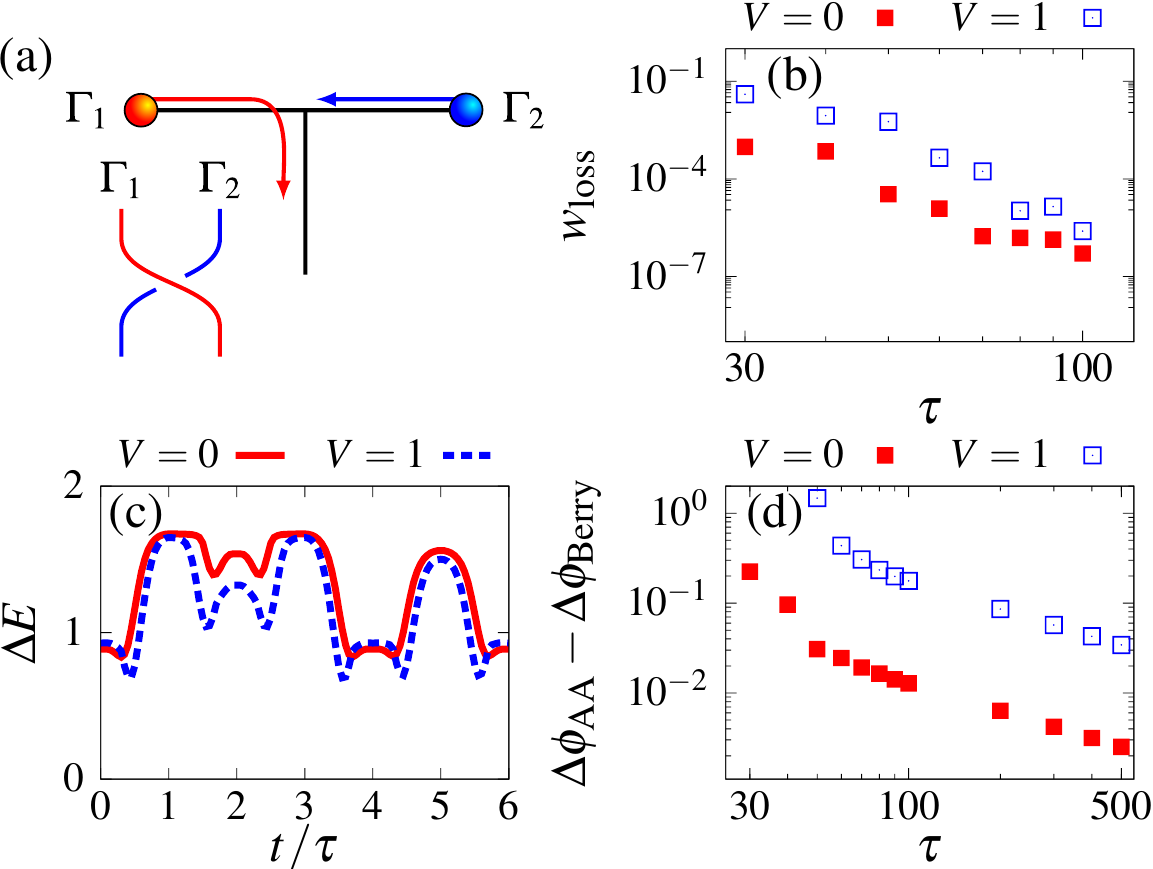}
    \caption{(a)  Sketch of trijunction hosting a pair of MZMs, $\Gamma_1$ and $\Gamma_2$, as well as the braiding procedure marked schematically with arrows;
    (b) loss of the fidelity, $w_\mathrm{loss}$, as a function of the total evolution-time $T=6\tau$;
    (c) energy gap $\Delta E$  vs. time $t/\tau$;
    (d) difference between $\Delta\phi_\AA$ and $\Delta\phi_\Berry$ vs. the evolution time $T=6\tau$.
    Results for system with ($V=1$) and without ($V=0$) many-body interactions (see labels) for $L=7$, $\Delta=0.8$, $\mu=0$.
    }
    \label{fig.fig1}
\end{figure}

We study the dynamics of a single qubit (sparsely) encoded in two pairs of MZMs, $\Gamma_1$, $\Gamma_2$ and  $\Gamma_3$, $\Gamma_4$. 
Each pair of MZMs  is hosted on a trijunction
schematically shown in Fig. \ref{fig.fig1}(a). 
The basis of the qubit consists of two states with even  total number of fermions, $|0 \rangle=|e_{12} \rangle \otimes |e_{34}\rangle$ and $|1 \rangle=|o_{12} \rangle \otimes |o_{34}\rangle$.
Here, $|e_{12} \rangle $ and $|o_{12} \rangle $ denote the states of the junction $J_{12}$ with even and odd number of fermions, respectively.
Similar notation holds for junction $J_{34}$.
 
We study the simplest setup which allows for the braiding of MZMs ~\cite{alicea.oreg.11,sekania.plugge.17}.
Namely, we consider a trijunction [cf. Fig. \ref{fig.fig1}(a)] consisting of  three chains of equal length and we set for each chain different  phase of the superconducting order parameter,  $\Delta_{ij} = \Delta \exp(-\mathrm i\varphi_{ij})$, where $\varphi_{ij}=0,\,+\tfrac\pi2,\,-\tfrac\pi2$ in left, right and the vertical chain, respectively. 
We assume also that each junction contains $L$ sites and is described by the Kitaev model~\cite{kitaev.01} with many-body interaction~\cite{thomale.rachel.13,wieckowski.maska.18,stoudenmire.alicea.11,hassler.schuricht.12,peng.pientka.15,gergs.niklas.16,dominguez.cayao.17},
\begin{eqnarray}
H(t) &=& H_0+\sum_{i} \mu_i(t) \widetilde n_i,  \label{eq.ham}  \\
H_0 &=& \sum_{\langle i,j\rangle}
\left[ ( t_0^{\phantom \dagger} a_i^{\dagger} a_{j}^{\phantom \dagger} + \Delta_{ij}^{\phantom{\dagger}}  a_i^{\dagger} a_{j}^{\dagger}) + \text{H.c.} \; + \; V\;  \widetilde n_i \widetilde n_{j}\right] \nonumber.
\end{eqnarray}
Here  $a_i^\dagger$ creates a fermion on site $i$,  $\widetilde n_i = a_i^\dagger a_i^{\phantom\dagger}-\tfrac 12$, $t_0$ is the hopping between the neighboring sites on a junction, $V$ is the nearest neighbor repulsion. 
The time-dependence of $\mu_i(t)$ allows to implement the braiding of MZMs as it is described below in more details.
We use dimensionless units, $\hbar = 1$ and $t_0=1$. 
 
In the case of a single Kitaev chain with a uniform and time-independent  $\mu_i(t)=\mu$,  one may switch between  the trivial and the topological phases via tuning the chemical potential.
In a system without many-body interaction ($V=0$) and non-zero $|\Delta|>0$, the topological phase is present for $|\mu|\le 2t_0$, while the trivial one for $|\mu|>2t_0$~\cite{kitaev.01,alicea.12}.  
The topological regime in a system with many-body interaction has been discussed, e.g., in~\cite{wieckowski.maska.18,thomale.rachel.13,katsura.schuricht.15}. The braiding is achieved via slow tuning of $\mu_i(t)$  in such a way that selected sites remain in topological regime whereas the other remain in the trivial regime~\cite{alicea.oreg.11}. 
Namely,
\begin{equation}
\mu_i(t) = \mu_c g_i(t)+\mu,
\label{prot}
\end{equation}
where  $\mu$ is the uniform chemical potential and we set $\mu_c=\pm4$. The details 
of the ramping protocol,  $ g_i(t) \in [0,1] $,  are the same as in Ref. \cite{sekania.plugge.17} and are recalled in the Appendix \ref{sec.protocol}.

The braiding protocol describes a cyclic evolution of the Hamiltonian \eqref{eq.ham} in the parameter space.
The many-body wave function is obtained from the numerical  solution ~\cite{fehske.schleede.09,vega.93} of the time-dependent Schr\"odinger equation $i \partial_t |\psi(t)\rangle={H}(t)|\psi(t)\rangle$. 
Initially ($t=0$), we set $\mu_i=\mu_c$ for sites $i$ in the vertical chain [cf. Fig. \ref{fig.fig1}(a)] which is then in the trivial regime.
Two remaining (horizontal) chains are in topological regime  and host two MZMs located at the edges of these wires. Next, by adiabatic tuning of $g_i(t)$, we control the boundaries of topological regime and swap the positions of $\Gamma_1$ and $\Gamma_2$, cf. Fig.~\ref{fig.fig1}(a). 
We split our protocol into six equal time-windows:
$(0,\tau)$ -- moving $\Gamma_1$ to the center of the  junction;
$(\tau,2 \tau)$ --  moving $\Gamma_1$ to the edge of vertical chain;
$(2\tau,3 \tau)$ -- moving $\Gamma_2$  to the center of the  junction;
$(3\tau,4 \tau)$ -- moving $\Gamma_2$ to the edge of the left chain;
$(4\tau,5 \tau)$ -- moving $\Gamma_1$ to the center of the  junction;
$(5\tau,6 \tau)$ --  moving $\Gamma_1$  to the edge of the right chain.
These steps are shown explicitly in  Fig. \ref{fig.sm0} presented in the in the Appendix \ref{sec.protocol}.

\section{Results}\label{sec.results}
\subsection{Geometric phase for a single trijunction} 
We examine the non-Abelian properties of MZMs by calculating the geometric phases: 
the Berry phase, $\phi_{\Berry}$,  in the case of the adiabatic evolution  \cite{berry.84} or the Aharonov--Anandan phase, $\phi_{\AA}$ in the case of  a general cyclic evolution \cite{aharonov.anandan.87}.
However first, we check when the evolution is cyclic, i.e., when the final quantum state $\left| \psi(T) \right> \left< \psi(T)\right|$ equals the initial one $\left| \psi(0) \right>\left< \psi(0)\right|$, where for the present protocol $T=6 \tau$.
 We examine this property by  calculating the loss of  the fidelity, 
\begin{equation}
\label{eq.fidelity}
w_{\mathrm{loss}} = 1 - \left|\left< \psi(T)|\psi(0) \right> \right|^2,
\end{equation}
which is shown in Fig.~\ref{fig.fig1}(b).  
One may observe that this quantity decreases  when the evolution-time  increases and becomes negligible for  $\tau \gtrapprox 100$. 
The necessary condition for the  adiabaticity of the time-evolution is  a non-vanishing energy gap between the ground-state and the first excited state.
 In Fig.~\ref{fig.fig1}(c) we show the  gap $\Delta E = \min\{E_1^o-E_0^o,\,E_1^e-E_0^e \}$ during the entire evolution, where $E_n^{e(o)} $ is the energy of the $n$-eigenstate in the even (odd) parity sector.
 Since $\Delta E$ does not vanish,  the evolution should be adiabatic for sufficiently large  $\tau$.
 
In the case a cyclic evolution the initial and the final  wave functions  differ only by the phase factor,
\begin{equation}
\label{eq.phaseDef}
    \left| \psi(T) \right> = e^{i\phi} \left| \psi(0) \right>= e^{i(\phi_{\dyn}+\phi_{\geo})} \left| \psi(0) \right>,
\end{equation}
which contains both the gauge-invariant  geometric  phase $\phi_{\geo}$ and the dynamical phase $\phi_{\dyn}$~\cite{anandan.christian.97}.
We evaluate the geometric phase from the standard expression \cite{mukunda.simon.93},
\begin{equation}
\label{eq.geometricPhase}
\phi_{\geo} = \arg(\left< \psi(0)|\psi(T) \right> ) - \arg\left( \prod_{j=0}^{N-1}  \left< \psi(t_j)|\psi(t_{j+1}) \right> \right),
\end{equation}
where $t_0=0$ and  $t_N=T$.
In the case of a generic cyclic quantum evolution, $\phi_{\geo}$ evaluated from Eq.~\eqref{eq.geometricPhase} represents $\phi_{\AA}$. 
Then, the wave function $\left|\psi(t_j) \right>$ is obtained directly from the time-dependent Schr\"odinger equation. 
In the case of the adiabatic cyclic evolution, $\phi_{\AA}= \phi_{\Berry}$.
Then, the wave functions $\left|\psi(t_j) \right>$ are obtained from diagonalization of the  instantaneous Hamiltonians $H(t_j)$. 

The essential quantity for implementing the Majorana quantum gates is  the difference between phases acquired during  evolution in sectors with even and odd particle numbers:
 $\Delta\phi_{\geo}=\phi^e_{\geo}-\phi^o_{\geo}$, (cf. the basis states of the qubit). 
As a final test of the adiabatic evolution, in  Fig.~\ref{fig.fig1}(d) we show that the  difference  $\Delta\phi_{\AA} - \Delta\phi_{\Berry}$ 
 decreases  with increasing  $\tau$.  
The latter difference is significantly larger for systems with many-body interactions, nevertheless one may expect that it vanishes
 for $\tau \rightarrow \infty$ also for $V\ne 0$. 
Therefore, from now on we focus only on the adiabatic evolution.    

Finally, we introduce the dynamical phase for the adiabatic evolution in the ground-states, e.g. for $|e_{12}(t)\rangle$:
\begin{equation}
    \phi_{\dyn}^{e,J_{12}} = \int\limits_0^T \mathrm dt\, \langle e_{12}(t)|H(t)|e_{12}(t)\rangle,\label{eq.dyn}
\end{equation}
and the phase-difference between even and odd parity sectors, 
\begin{equation}
\Delta\phi_{\dyn}^{J_{12}}=\phi_{\dyn}^{e,J_{12}}-\phi_{\dyn}^{o,J_{12}}.
\end{equation}
The latter quantity is proportional to the difference of the ground-state energies in various sectors,  $\delta E=E_0^e-E_0^o$.
One tries to eliminate the dynamical phase and work only with the geometric phase.

\begin{figure}[t!]
    \centering
    \includegraphics[width=\columnwidth]{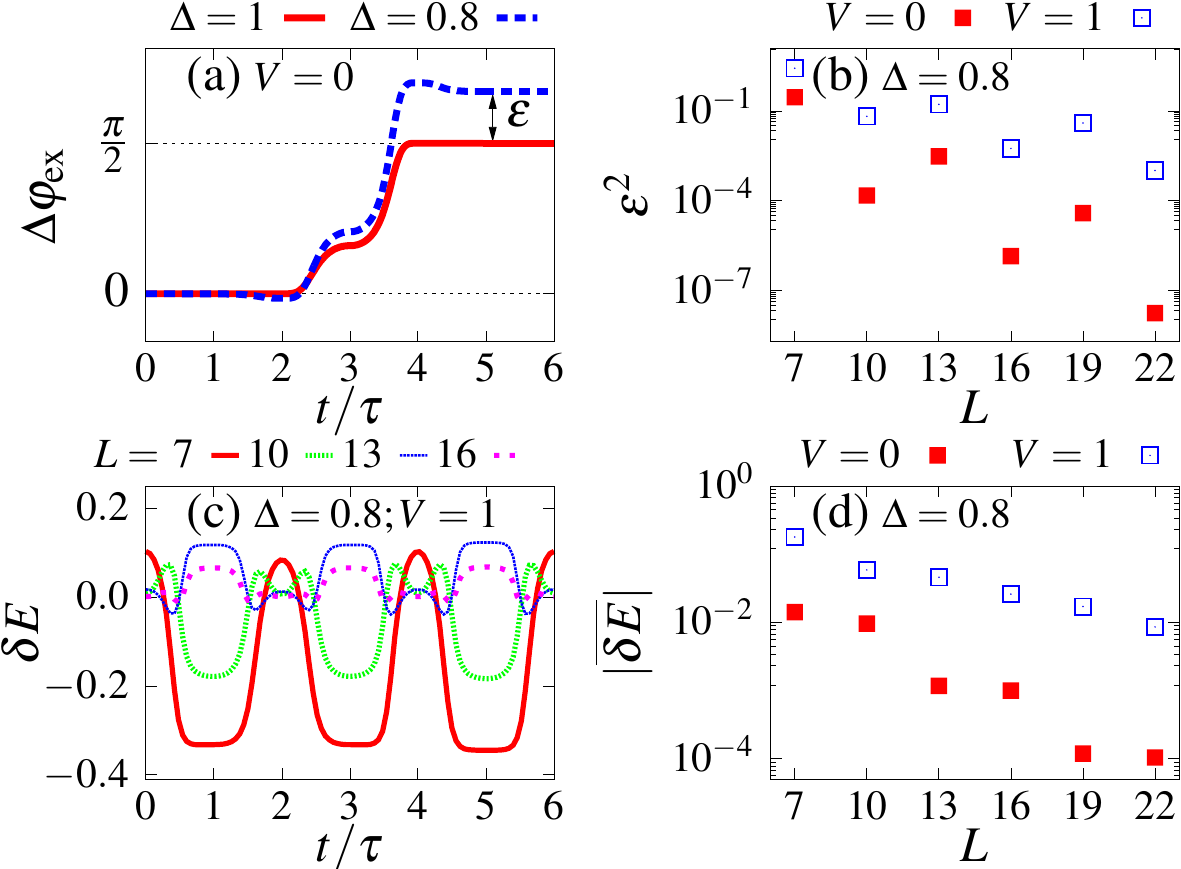}
    \caption{
    Single braiding on a single trijunction for $\mu=0$:
    (a) exchange phase $\Delta\varphi_\ex$ during evolution as a function of time $t/\tau$ for $L=7$;
    (b) finite-size scaling of the braiding error $\epsilon$;
    (c)  split between the instantaneous ground-states energies in different parity sectors  $\delta E(t)=E_0^e(t)-E_0^o(t)$;
    (d) finite-size scaling of the average energy split $\delta E$. 
    }
    \label{fig.fig2}
\end{figure}

If Majorana fermions are separated  in the real-space then they are strict zero-modes, hence  $E_0^e=E_0^o$ and $\Delta\phi_{\dyn}^{J_{12}} = 0$. Then, 
the only contribution to the phase difference comes from the geometric phase, $\Delta \phi=\Delta\phi_\Berry$. 
It is well established  that the braiding of strict MZMs leads to $\Delta\phi_\Berry=\pm\frac\pi2$ \cite{alicea.oreg.11}.
In order to construct the phase-gate based on the geometric-phase, one needs a protocol for which $\Delta\phi_\Berry = \pi/4$~\cite{nielsen.chuang.11}, whereas
  $\Delta\phi_{\dyn}=0$.  
  Such gate is unprotected by topology but it is adiabatic, i.e., the acquired phase is independent of the  evolution time. 
  
\subsection{Braiding error due to overlap of the Majorana fermions}
The geometric phase is  a gauge-invariant quantity provided that the Hamiltonian follows a closed loop in the parameter space.
However, in order to gain more insight, we introduce also the adiabatic \textit{exchange phase}  $\Delta\varphi_\ex(t)$, defined for arbitrary $0 \le t \le T$~\cite{sekania.plugge.17},
such that $\Delta\phi_\Berry=\Delta\varphi_\ex(T)$.  Fig.~\ref{fig.fig2}(a) shows the latter quantity. 
For a special case  $\Delta=1$, the MZMs are located on single edge sites   and do not overlap during the braiding even for a finite system.  
Then, $\Delta\phi_\Berry= \tfrac \pi2$ for arbitrary $L$.
However, for $\Delta\neq1$ and finite $L$, such phase  deviates from $\tfrac\pi2$ by a braiding error, $\epsilon=\Delta\phi_{\Berry}-\tfrac \pi 2$. Fig.~\ref{fig.fig2}(b)  demonstrates that the braiding error is  a finite-size effect. 
In the case of an infinite trijunction, when the MZMs are fully separated in the real space, $\epsilon$ seems to vanish and the Berry phase  equals $\tfrac\pi2$.

We stress that a non-zero braiding error is intimately connected with a non-vanishing dynamical phase. 
Overlap of the Majorana fermions  lifts the degeneracy of the
ground-state, $E_0^e \ne E_0^o$, hence in general  $\Delta\phi_{\dyn}^{J_{12}} \ne 0$.  
The energy splitting, $\delta E $,  depends  on the distance between the Majorana fermions which varies during the evolution.  
In Fig.~\ref{fig.fig2}(c) we show the instanteneous  $\delta E$ as a function of the evolution time $t/\tau$ for different system sizes $L$.  
It is clear that $\delta E$
decreases when $L$ increases. In order to discuss this effect in more details, we have calculated the average splitting $\overline{\delta E}=\frac1T\int_o^T \mathrm dt\,\delta E(t)$
which determines also the dynamical phase $\Delta\phi_{\dyn}^{J_{12}}= T \overline{\delta E}$. 
Fig.~\ref{fig.fig2}(d) shows the finite-size scaling of $\overline{\delta E}$ which seems to decay  almost exponentially with  increasing $L$. 

\begin{figure}[!t]
    \centering
    \includegraphics[width=\columnwidth]{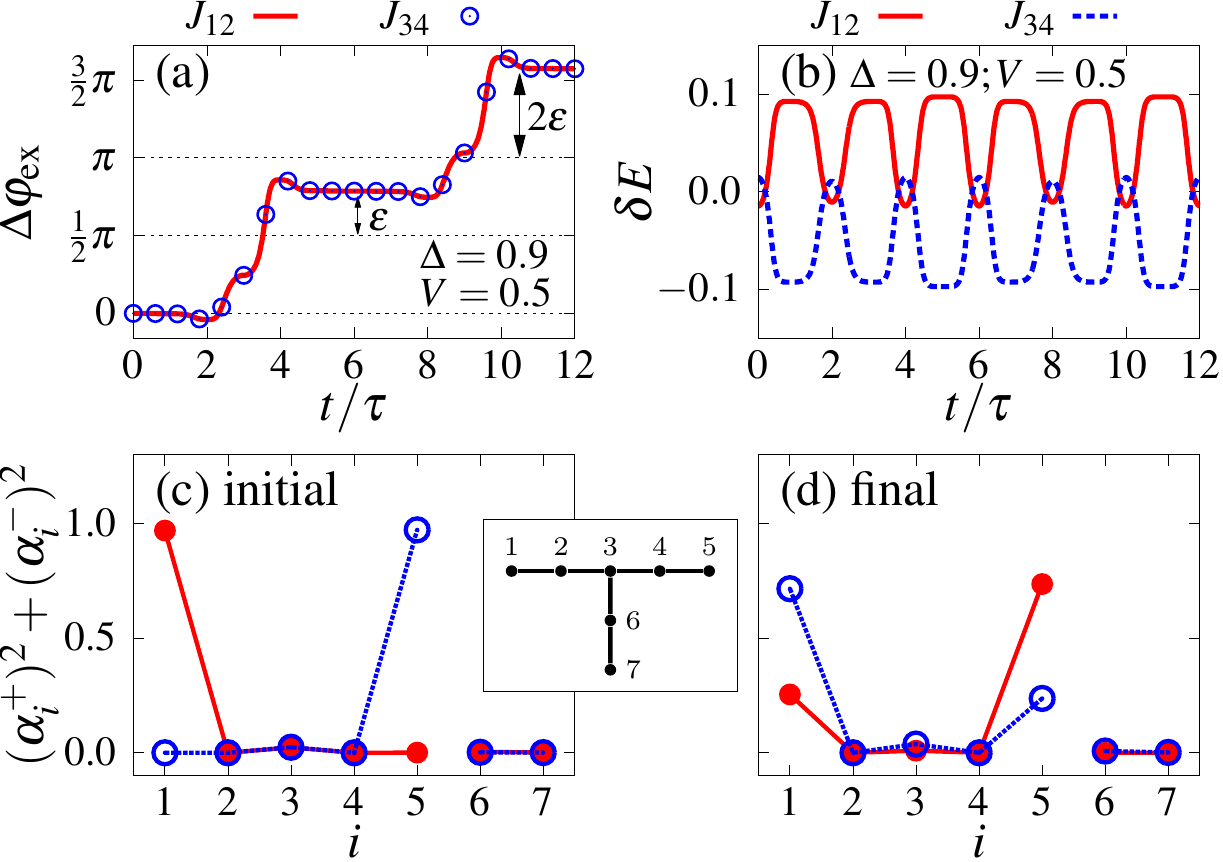}
    \caption{
    (a)--(b): Double braiding  for $\mu=0$. We set $\mu_c=4$ and $\mu_c=-4$ for the  junction $J_{12}$ and $J_{34}$, respectively.
    (a) exchange phase $\Delta\varphi_\ex$  and  (b) energy splittings $\delta E$  determined separately  for each junction.
    Spatial structures of Majorana fermion $\Gamma_1$ (red solid line) and $\Gamma_2$ (blue dashed line) (c) before and (d) after single braiding for $\Delta=0.8$  and $V=0$.
     ($L=7$, $\mu=0$) 
    }
    \label{fig.fig3}
    
\end{figure}

\subsection{Cancelation of  the dynamical phases}
While the braiding error should be avoided in the topologically protected operations, it may still be very useful  for constructing the phase-gate with arbitrary phase-shift. 
The advantage of such solution over the simplest protocol based on getting Majorana fermions close to each other, consists in that  $\Delta\phi_\Berry$ does not depend on the total evolution time $T$.
However, a non-zero braiding error is intimately connected with a nonzero dynamical phase. 
Therefore, the idea of using the Berry phase 
would be useless unless one finds a method of eliminating the dynamical phase.
Below we show that $\Delta\phi_{\dyn}$ can indeed be eliminated by appropriate tuning of  junctions  which build the Majorana qubit.

We assume that both trijunctions, $J_{12}$ and $J_{34}$,  are described by the same Hamiltonian~\eqref{eq.ham} which is particle--hole symmetric up to the term containing $\mu_i$, see Eq.~\eqref{prot}.
Each junction contains  {\it odd} number of sites and  the braiding is applied {\it twice} to each junction. 
However, one applies positive $\mu_i$ for one junction and a negative   $\mu_i$ for the other. Namely,
the trijunctions $J_{12}$ and $J_{34}$ are described, respectively, by the Hamiltonians 
\begin{eqnarray}
H_{12} (\Delta_{ij})  & =  & H_0 (\Delta_{ij}) + \sum_{i} \mu_i(t) \widetilde n_i, \\
H_{34} (\Delta_{ij}) & =  & H_0 (\Delta_{ij}) - \sum_{i} \mu_i(t) \widetilde n_i,
\end{eqnarray}
where for clarity of the present discussion we explicitly mark the dependence of Hamiltonians on the  superconducting order parameter. 

In Fig.~\ref{fig.fig3}(a) we present the geometric phase,  $\Delta\varphi_\ex$,  gained by each junction during such double-braiding protocol.
The sign of $\mu_i$ does not influence the geometric phase and we find $\Delta\phi^{J_{12}}_\Berry= \Delta\phi^{J_{34}}_\Berry= \pi+2\epsilon$.
However, Fig.~\ref{fig.fig3}(b) shows that the energy splittings $\delta E$  for the trijunctions $J_{12}$ and $J_{34}$ have opposite signs, hence $\Delta\phi_{\dyn}^{J_{12}}+\Delta\phi_{\dyn}^{J_{34}}=0$.  
In order to explain the latter identity we assume that  the sites within each junction are enumerated according to the scheme shown in the inset in Fig.  \ref{fig.fig3}(c).
Then, it is easy to check (for odd $L$)  that the  neighboring  sites  $\langle i,j \rangle$ are labeled by integers
with opposite parities, i.e.,  if $i$ is odd then $j$ is even. 
In other words, the trijunctions form a bipartite lattices consisting of two sublattices which contain, respectively, odd and even lattice sites $i$.  

We consider a standard  particle--hole (Shiba) transformation~\cite{essler.frahm.05},
\begin{equation}
U=(a^{\dagger}_L-a_L)(a^{\dagger}_{L-1}+a_{L-1}) \; ... \; (a^{\dagger}_2+a_2)(a^{\dagger}_1-a_1),
\end{equation}  
for which $U^{\dagger} U =  U U^{\dagger} =1$ and  $U a_i U^{\dagger}=(-1)^i a^{\dagger}_i$.  
One finds that Hamiltonians of both junctions are connected via this particle--hole transformation
\begin{equation}
U  \: H_{12} (\Delta_{ij}) \; U^{\dagger} = H_{34} (\Delta^{*}_{ij}),
\end{equation}
whereas the parity operator
\begin{equation}
P=\prod_{i=1}^{L}(1-2 a^{\dagger}_i a_i)
\end{equation} 
is odd under the latter transformation, $U  P U^{\dagger}= (-1)^L P=-P$.  Considering an eigenstate   $ | n \rangle $ of $H_{12} (\Delta_{ij})$, 
\begin{equation}
 H_{12} (\Delta_{ij}) | n \rangle =E_n  | n \rangle, \quad   P  | n \rangle = p_n  | n \rangle,
 \end{equation} 
 one finds that $U |n \rangle $ is  an eigenstate of $H_{34} (\Delta^{*}_{ij})$ with energy $E_n$ but with the opposite parity,
$ P  U  | n \rangle=  -p_n  U  | n \rangle$.  Therefore,  $H_{12} (\Delta_{ij})$  and $H_{34} (\Delta^{*}_{ij})$ have the same energy spectra, however, with swapped parities of the energy levels.
It is also clear that the energy spectrum of $H_{34} (\Delta^{*}_{ij})$ is the same as that of  $H_{34} (\Delta_{ij})$.

Overlap of the Majorana fermions lifts the ground state degeneracy, however the above particle--hole transformation holds true during the entire quantum evolutions. Then, using Eq.~\eqref{eq.dyn}  one finds that
$ \phi_{\dyn}^{e,J_{12}} = \phi_{\dyn}^{o,J_{34}}$ as well as $ \phi_{\dyn}^{o,J_{12}} = \phi_{\dyn}^{e,J_{34}}$ and, consequently, 
\begin{equation}
\Delta\phi_{\dyn}^{J_{12}} + \Delta\phi_{\dyn}^{J_{34}}=\phi_{\dyn}^{e,J_{12}}-\phi_{\dyn}^{o,J_{12}} +\phi_{\dyn}^{e,J_{34}}-\phi_{\dyn}^{o,J_{34}}=0. 
\end{equation}
After the double-braiding protocol, the initial state of the qubit  $|\psi(0) \rangle=  |0 \rangle +|1\rangle$  will become   
\begin{equation}
 |\psi(2T) \rangle =  \exp(i \chi) (|0 \rangle +\exp( 4 i \epsilon) |1\rangle),
 \end{equation}
 hence the relative phase, $4 \epsilon$, is determined solely   by the braiding error for the geometric phase.  

\subsection{Tuning of the  geometric phase}

\begin{figure}[t]
    \centering
    \includegraphics[width=\columnwidth]{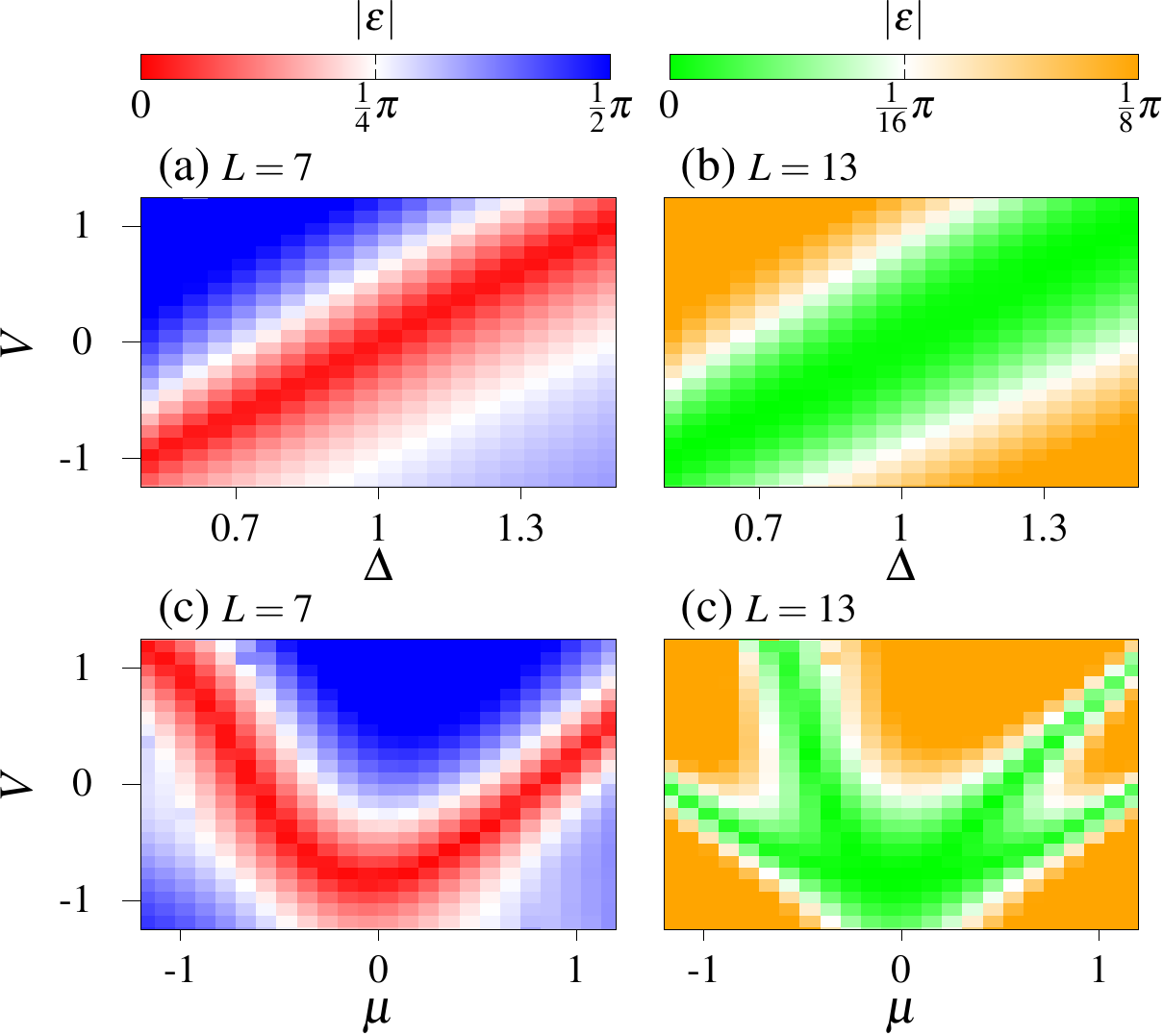}
    \caption{(color online)  Absolute value of the braiding error $\epsilon$ (i.e., the deviation of the Berry phase $\Delta\phi_{\Berry}$ from $\tfrac \pi 2$):
    (a)--(b) $|\epsilon|$ as a function of interaction $V$ and superconducting gap $\Delta$ ($\mu=0$), for $L=7$ and $L=13$, respectively;
    (c)--(d) $|\epsilon|$ as a function of interaction $V$ and chemical potential $\mu$ ($\Delta=0.6$) for $L=7$ and $L=13$, respectively.
    }
    \label{fig.sm3}
\end{figure}

The main result of the present work concerns the phase-gate based on the braiding error $\epsilon$, i.e., the deviation of the Berry phase $\Delta\phi_{\Berry}$ from $\tfrac \pi 2$. The braiding error vanishes for the topologically protected gates when one braids non-overlapping MZMs on an infinite trijunction.
However in order to construct the standard $\pi/8$-gate, one needs $\epsilon=\pi/16$.
Here we show for a finite junction that $\epsilon$ may be rather easily tuned via changing the parameters of the Hamiltonian \eqref{eq.ham}.

Since $\epsilon$ originates from the overlap of Majorana fermions, it strongly depends on the system size. Fig.~\ref{fig.sm3} shows $|\epsilon|$ for $L=7$ and $L=13$, which are the smallest system sizes with odd $L$. 
In Figs.~\ref{fig.sm3}(a)--\ref{fig.sm3}(b) we show how $|\epsilon|$ depends on the many-body interaction $V$ and the superconducting order parameter $\Delta$ for $\mu=0$. Figs.~\ref{fig.sm3}(c)--\ref{fig.sm3}(d) show the same quantity for $\Delta = 0.6$ as a function of  the many-body interaction $V$ and chemical potential $\mu$. Tuning the superconducting order parameter or the interaction strength
is (most probably) not relevant for realistic experimental setups. Therefore, the most important result is that $\varepsilon$ may be well tuned via changing  the chemical potential.

\subsection{Spatial structure of overlapping Majorana fermions after braiding}
In order to follow the spatial structure of the overlapping Majorana fermions  ($\Gamma_1$ and $\Gamma_2$) during a single adiabatic braiding on a single junction $J_{12}$, 
we represent both fermions as a linear combination of  the local Majorana operators $\gamma^+_i=a_i+a_i^\dagger$ and $\gamma^-_i=i(a_i-a_i^\dagger)$, namely
 $\Gamma_m =\sum_i^L (\alpha_{i}^{m,+} \gamma_i^++\alpha_i^{m,-}\gamma_i^-)$ for $m\in \{1,2\}$. Then, we apply the algorithm developed in Ref. \cite{wieckowski.maska.18} to find the coefficients, $\alpha_{i}^{m,\pm}$ 
for each instantaneous  Hamiltonian $H(t)$. This algorithm targets the MZMs following their formal definition via the commutation relations ~\cite{sarma.freedman.15}:  $\{\Gamma_m, \Gamma_{m'} \}=2\delta_{m,m'}$ and $[\Gamma_m,H]=0$. The latter commutation relations  are invariant under the rotation $\vec{\Gamma} \rightarrow  \mathcal O(\beta) \vec{\Gamma}$, where 
\begin{equation}
\vec\Gamma=
    \left(\begin{array}{c}
    \Gamma_1\\
    \Gamma_2
    \end{array}
    \right),\quad \quad 
\mathcal O(\beta) = \left(\begin{array}{cc}
         \cos\beta & -\sin\beta\\
         \sin\beta &\phantom+\cos\beta
    \end{array}\right),
    \end{equation}
  hence  also the  coefficients, $\alpha_{i}^{m,\pm}$,  are defined up to  an arbitrary choice of $\beta$.  
Initially at time $t=t_0$, we choose the angle $\beta(t_0)$ following the standard convention in that $\Gamma_1$ and $\Gamma_2$ are located at the opposite edges of trijunction, as shown in Fig.~\ref{fig.fig3}(c).
Then, for each time $t_j$ during the adiabatic evolution, we find $\beta(t_j)$ that minimizes the (squared) distance
\begin{equation}
    \left\|
    \vec\Gamma(t_{j})
    -
    \vec\Gamma(t_{j-1})
    \right\|^2 =\sum_{i=1}^{L}\sum_{m=1}^{2}\sum_{s=\pm}  [\alpha_{i}^{m,s}(t_{j})- \alpha_{i}^{m,s}(t_{j-1})]^2.
\end{equation}
If Majorana fermions are strict zero-modes  then this approach reproduces the standard braiding that swaps the MZMs, i.e., $\Gamma_1(T)= \pm \Gamma_2(0)$ and $\Gamma_2(T)  = \mp \Gamma_1(0)$.  The latter swapping may also be written as   
$ \vec\Gamma(T)=\mathcal O(\Delta\phi_\Berry)  \vec \Gamma(0)$ for $\Delta\phi_\Berry=\pm\frac\pi2$.   
It turns out, that the latter relation holds true also for $\Delta\phi_\Berry \ne \pm\frac\pi2$, i.e., also for braiding of the overlapping Majorana fermions. 
Then however, the cyclic evolution cannot be understood as simple swapping of the Majorana fermions.
In particular $\Gamma_1(T)$ becomes a linear combination of both  $\Gamma_1(0)$ and $\Gamma_2(0)$, hence it contains non-vanishing contributions located at both edges of the junction, see  Figs.~\ref{fig.fig3}(c)--\ref{fig.fig3}(d). The latter holds true whenever $\Delta\phi_\Berry$  is not a multiple of $\pm\frac\pi2$.

\subsection{Phase-gate constructed from the braiding error}

\begin{figure}
    \centering
    \includegraphics[width=\columnwidth]{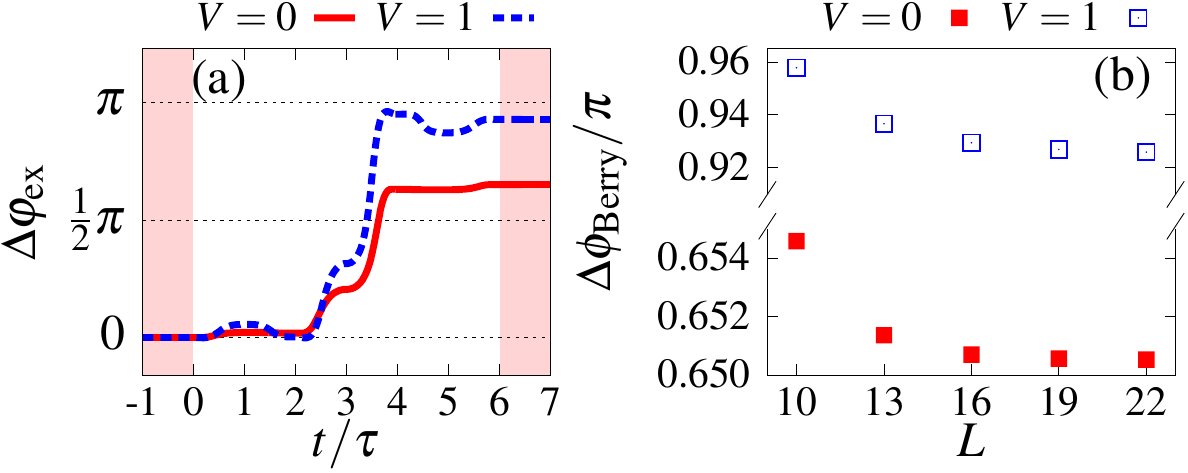}
    \caption{
    Braiding of MZMz which are brought together for $t \in (-\tau,0)$ and shifted apart for $t\in(6 \tau,7\tau)$. We set $\Delta=0.8$, $\mu=0$.
    (a) exchange phase $\Delta \varphi_\ex$ for $L=19$ and $L'=7$.  
    (b) Geometric phase $\Delta\phi_\Berry$ for $L'=7$ vs. $L$.
    }
    \label{fig.fig4}
\end{figure}

It is desirable to have a single junction on which one may perform the topologically protected operations, e.g., braiding of separated MZMs with $\Delta\phi_\Berry= \pm\frac\pi2 $,
but also the unprotected  adiabatic operations, e.g., the braiding of overlapping MZMs with $\Delta\phi_\Berry \ne \pm\frac\pi2$. 
Due to the former operations, the trijunction with $L$ sites should be as large as possible. 
Then, in order to perform the also latter operation, one needs to bring both MZMs towards the center of the junction,  so they start to overlap. 
This may be achieved via appropriate tuning of  $\mu_i(t)$ in the time-window $t \in (-\tau,0)$, see the left shaded area in 
Fig.~\ref{fig.fig4}(a).
Then, one may carry out the braiding protocol on the  restricted  trijunction with $L' \ll L $ sites in the time-window $t\in(0, 6 \tau)$. 
Finally, for $t\in(6 \tau,7\tau)$  the Majorana fermions are shifted apart to their original positions at the edges of the unrestricted (infinite) junction with $L$ sites, see the right shaded area in  Fig.~\ref{fig.fig4}(a).  

Fig.~\ref{fig.fig4}(a)  shows the exchange phase $\Delta\varphi_\ex(t)$ for $L=19$ and $L'=7$. 
We note rather negligible changes of   $\Delta\varphi_\ex(t)$  during the time-windows when MZMs are brought together, 
$(-\tau,0)$, or when they are shifted apart,   $(6 \tau,7\tau)$.
Fig.~\ref{fig.fig4}(b) shows one of main results of the present work: the finite-size scaling of the geometric phase $\Delta\phi_{\Berry}$ for  fixed $L'=7$ and various $L$.
In contrast to results in Fig.~\ref{fig.fig2}(a), the braiding error is not a finite-size effect and remains non-zero also for $L\rightarrow \infty$ provided  that $L'$ is finite and Majorana fermions
overlap during the braiding protocol.
Weak $L$-dependence of $\Delta\phi_{\Berry}$  in Fig. \ref{fig.fig4}b may originate from 
the leakage of MZMs   into these sites which remain in the trivial regime ~\cite{klinovaja.loss.12,ruiz-tijerina.vernek.15,ptok.kobialka.17}.

\section{Conclussions}\label{sec.summary}
We have studied  the dynamics of a qubit built out of four Majorana quasiparticles evolving on two trijunctions. We  focused on a case when Majorana fermions evolve on spatially restricted junctions.
Due to their mutual overlapping, they are not strict zero-modes anymore, hence the qubit acquires both dynamical and geometric phases during the braiding protocol.   
We have demonstrated that the dynamical contribution may be cancelled out  if the trijunctions are described by the same particle--hole symmetric Hamiltonian and contain odd number of the lattice sites.
The geometric contribution deviates from that for braiding of strict zero-modes  $\Delta\phi_\Berry= \pm\frac\pi2 $,  and the latter deviation allows one to build the adiabatic phase-gate with tunable phase-shift. The only difference with respect to the topologically protected braiding of MZMs consists in that the Majorana fermions are brought together before the braiding and  are shifted apart  after the braiding. 
Probably, the protocol still should be followed by some error correction, however the initial error is expected to be smaller than in standard protocol based on the dynamical phase.

\begin{acknowledgments}
We acknowledge fruitful discussions with A. Ptok and M. Maśka.
This work was supported by the National Science Centre, Poland, under Grant No. 2016/23/B/ST3/00647.
\end{acknowledgments}

\begin{appendix}

\section{Smooth ramping protocol} 
\label{sec.protocol}

\begin{figure}[b]
    \centering
\includegraphics[width=1.\columnwidth]{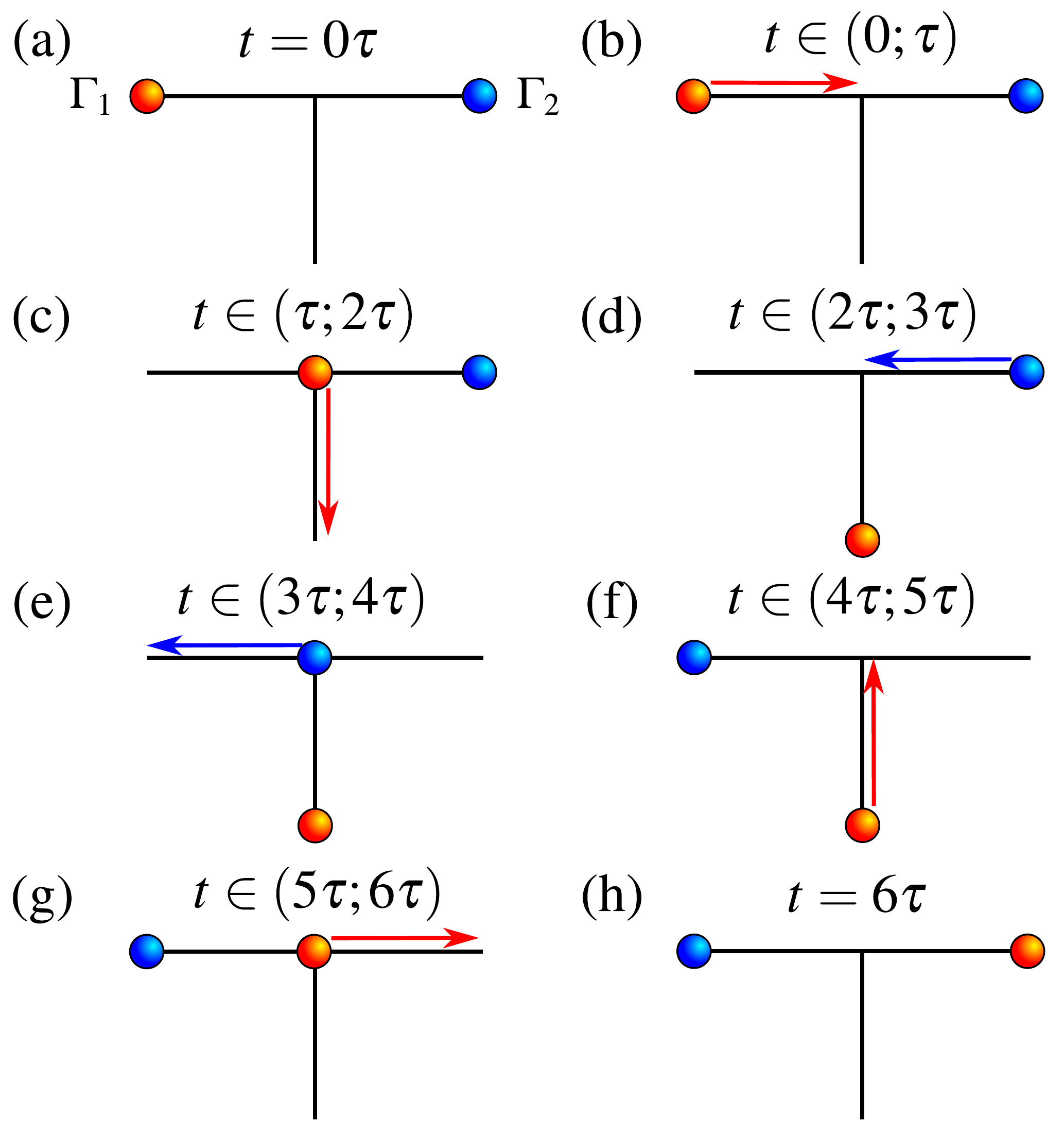}
\caption{
Braiding protocol:
(a) initial position of MZMs;
(b) moving $\Gamma_1$ to the center of the  junction;
(c) moving $\Gamma_1$ to the edge of vertical chain;
(d) moving $\Gamma_2$  to the center of the  junction;
(e) moving $\Gamma_2$ to the edge of the left chain;
(f) moving $\Gamma_1$ to the center of the  junction;
(g) moving $\Gamma_1$  to the edge of the right chain;
(h) final position of MZMs.
}
\label{fig.sm0}
\end{figure}

\begin{SCfigure*}
    \centering
    \includegraphics[width=1.5\columnwidth]{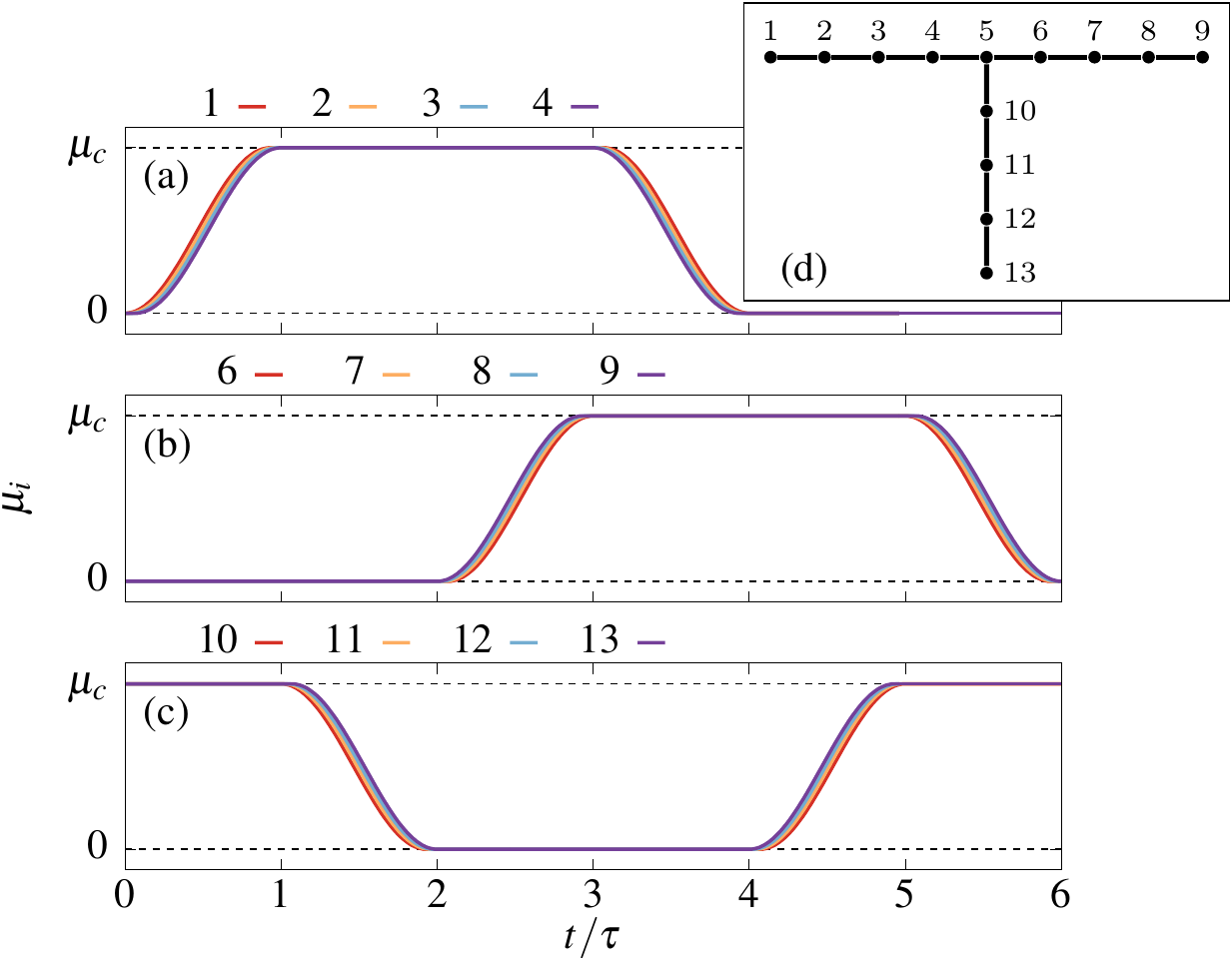}
    \caption{
    Standard braiding protocol.
    (a)--(c) potentials $\mu_i$ as a function of time $t/\tau$ in: (a) left  (b) right and (c) vertical chain of the trijunction. The numbering of sites is shown in the panel (d).
    }
    \label{fig.sm1}
\end{SCfigure*}

\begin{SCfigure*}
    \centering
    \includegraphics[width=1.5\columnwidth]{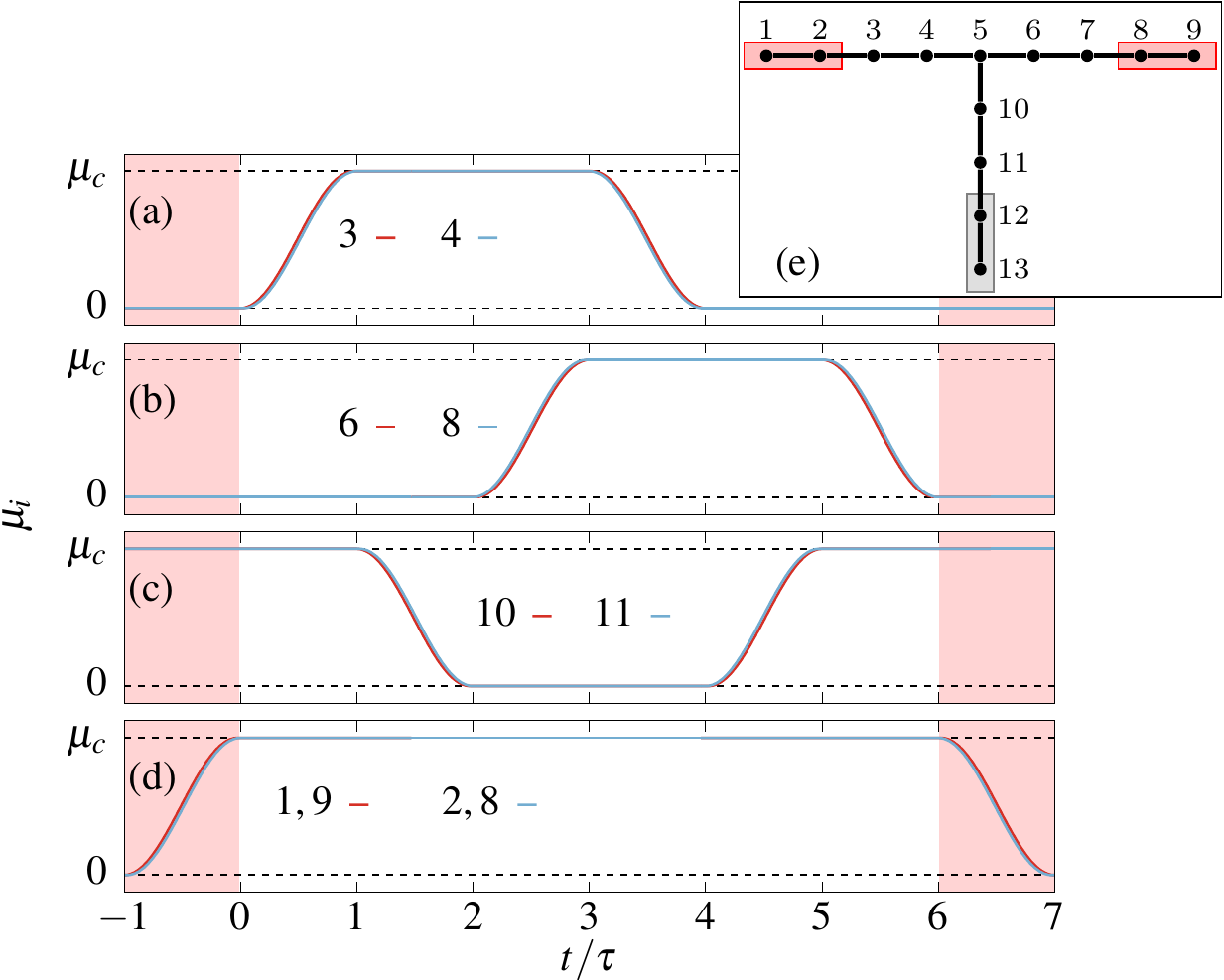}
    \caption{
    Extended braiding protocol in which MZMs are brought together for $t \in (-\tau,0)$ and shifted apart for $t\in(6 \tau,7\tau)$.
    (a)--(d) potentials $\mu_i$ as a function of time $t/\tau$ for selected sites of: (a) left  (b) right and (c) vertical chain of the trijunction, which is schematically represented in the panel (e); (d) shows
    $\mu_i$ for sites marked  with  red rectangles in (e).
    For sites which are marked with gray rectangles in the (e) we set $\mu_i(t)=\mu_c$ throughout the protocol.
    }
    \label{fig.sm2}
\end{SCfigure*}

We use exactly the same smooth ramping protocol as in Ref. \cite{sekania.plugge.17}.
Fig.~\ref{fig.sm0} illustrates subsequent steps of the braiding protocol and the corresponding time-windows.
The swap of MZMs is achieved via appropriate tuning of $\mu_i(t)$, see Eqs.~\eqref{eq.ham} and \eqref{prot} in the main text. 
Whenever we ramp-up selected sites, we use the following time-dependent function $g_i(t)\in[0,1]$:
\begin{equation}
g_i(t) = m\left(\frac{t}{\tau}[1+\alpha(\ell-1)] - \alpha(\ell-i)\right), \quad t\in[0,\tau],\label{eq.ramping}
\end{equation}
where we take $\alpha=0.025$ and $\ell$ is the length of each chain, i.e., $L=3\ell+1$.
Here,  $m(x)$ is a scalar function $
  m(x) = \sin^2\left( \frac{\pi}{2} r(x)\right)  
$ and $r(x)$ is linear ramp
$ r(x) = \min[\max(x,0),1]$.
For ramp-down protocol, we replace $t\to\tau-t$ in Eq.~\eqref{eq.ramping} to reverse the process in time.
Fig.~\ref{fig.sm1} shows $\mu_i(t)$ for the standard braiding protocol relevant for Fig.~\ref{fig.fig2}
in the main text. 
Fig.~\ref{fig.sm2} shows the same but for the extended protocol in which MZMs are first brought together for $t\in(-\tau,0)$, braided for $t\in(0,6\tau)$ and shifted apart for $t\in(6\tau,7\tau)$, see Fig.~\ref{fig.fig4} in the main text.

\sidecaptionvpos{figure}{c}


\end{appendix}

\clearpage

\bibliography{lib}

\end{document}